\documentclass[twocolumn,10pt]{article}

\title{An Estimation Method of a Constitutive-law for the Rod Model of DNA using
              Discrete-Structure Simulations}
\author{$^{\dag}$Adam R. Hinkle \and $^{\dag}$Sachin Goyal \and $^{*}$Harish J. Palanthandalam-Madapusi \vspace{0.25in} \\ \textit{$^{\dag}$Theoretical and Applied Mechanics, Cornell University, Ithaca, NY 14853-1503} \\ \textit{$^{*}$Mechanical and Aerospace Engineering, Syracuse University, Syracuse, NY 13244}}
\date{8 June 2009}

\usepackage{epsf,graphicx}
\usepackage{setspace,cite} 

\usepackage{rotating}

\usepackage{mathbbol}

\begin{document}
\maketitle

\begin{abstract}
The continuum-rod model has emerged as an efficient tool to describe the long-length-scale structural-deformations of DNA which are critical to understanding the nature of many biological processes such as gene expression. However, a significant challenge in continuum-mechanics-based modeling of DNA is to estimate its constitutive law, which follows from its interatomic bond-stiffness. Experiments and all-atom molecular dynamics (MD) simulations have suggested that the constitutive law is nonlinear and non-homogeneous (sequence-dependent) along the length of DNA. In this paper, we present an estimation method and a validation study using discrete-structure simulations. We consider a simple cantilever-rod with an artificially constructed, discrete lattice-structure which gives rise to a constitutive law. Large deformations are then simulated. An effective constitutive-law is estimated from these deformations using inverse methods. Finally, we test the estimated law by employing it in the continuum rod-model and comparing the simulation results with those of discrete-structure simulations under a different cantilever loading-conditions.
\end{abstract}

\section*{INTRODUCTION}

DNA is a long chain bio-polymer which contains the coded, genetic information needed to synthesize proteins and thus sustain all life~\cite{calladine:04a}. The biological functions of DNA such as gene expression are known to be significantly influenced by its long length-scale structural deformations such as looping~\cite{schleif:92a,semsey:05a}, which in turn is tied to its chemical make-up---the base-pair sequence. For example, the activity of the genes in lac-operon in the bacterium \emph{E.coli} is governed by the sequence-dependent looping behavior of its DNA segment adjacent to the genes (e.g. refer to~\cite{goyal:07a} and citations therein). In fact, this example has become a paradigm in understanding looping as a common gene-regulatory mechanism. How these long length-scale deformations depend on the discrete atomic structure molecule is still an open question. 

\begin{figure}[h!]
\centering
\includegraphics[width=85mm]{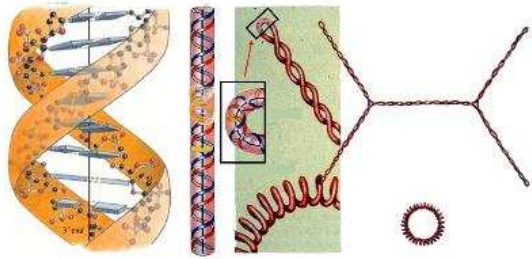}
\caption{Three different length scales of the DNA molecule. The smallest scale (left) shows double-helix structure (sugar-phosphate backbones (orange) and base-pairs (blue)). Intermediate scale (middle) shows several helical turns of double-stranded DNA (ds-DNA) and its helical axis (black). The largest scale (right) shows how the helical axis may curve to form supercoiling. (Courtesy: Branden and Tooze, 1999, From
Introduction to Protein Structure by Carl Branden \& John Tooze (reproduced by permission of Garland Science/Taylor \& Francis, LLC)~\cite{branden:99a} and Lehninger  et al. (copied with permission from W.H. Freeman)~\cite{lehninger:05a}).}
\label{fig:lengthscales}
\end{figure}

The DNA molecule itself is found in the nuclei of all living cells. It consists of two helices, which wind and join together by the hydrogen bonds between four bases: adenine (A), cytosine (C), guanine (G), and thymine (T). The only bonding which occurs is that of A with T (via double hydrogen bonds) and C with G (via triple hydrogen bonds)~\cite{watson:53a,franklin:53a}. Each base in DNA is also attached to a sugar molecule and a phosphate molecule. Together, a base, sugar, and phosphate are called a nucleotide. Nucleotides are arranged in two long strands that form the spiral called a double-helix. The structure of the double helix is somewhat like a ladder, with the base pairs forming the ladder's rungs and the sugar and phosphate molecules forming the vertical sidepieces of the ladder as in Figure~\ref{fig:lengthscales}.  The sequence of base pairs along the helix is the genetic code. 

Figure~\ref{fig:lengthscales} depicts DNA at different length scales. The familiar double-helix lies at the smallest scale with an approximate diameter of 2 nanometers (nm). A complete helical turn extends over nearly 3.6 nm. The center images of Figure~\ref{fig:lengthscales} show the length scale of a gene. This solid strand of DNA extends over tens to hundreds of helical turns (tens to hundreds of nanometers). The base-pair sequence within a gene constitutes a chemical code for the production of a specific protein via a process known as transcription. DNA also needs to make copies of itself via a process known as replication. This is critical when cells divide because each new cell needs to have an exact copy of the DNA present in the old cell. However, it is the looping (bending and twisting) at even longer length scales which critically influences these biological processes of replication and transcription. The right portion of Figure~\ref{fig:lengthscales} illustrates these longer-length scale structures which are highly curved and twisted topologies called supercoils. The supercoiling of DNA is also necessary to provide a means for these very long molecules to fit within the confines of cellular structures, such as the nucleus.

We are interested in assembling a computational framework which can successfully describe and predict how DNA undergoes these deformations. Among the attempts to model the structural mechanics of DNA molecules, few are simple and computationally efficient. One such successful approach to modeling mechanical behavior of DNA molecules is a continuum rod model~\cite{balaeff:06a,goyal:07a,goyal:08b,schlick:95a,olson:96a}, where elastic properties are prescribed and vary with length according to the local base-pair sequence of the molecule. However, a key component of the elastic rod model is the constitutive law (material properties), which is largely unknown. There is no clear consensus on its functional form and how it maps from the base-pair sequence. Although several experiments have provided the estimates of linear approximations of average torsional and bending stiffness by measuring persistence lengths~\cite{hagerman:88a,strick:96a,smith:08a,forth:08a}, recent experimental observations and model fitting suggest that the constitutive law is nonlinear~\cite{law,wiggins:05a,cloutier:04a,smith:08a,forth:08a}. The predictions of the continuum rod model are extremely sensitive to the constitutive law employed and its success hinges upon accurate modeling of the constitutive law.

Recently the development of an inverse formulation of the continuum rod model has been initiated to provide an estimation method of the constitutive law~\cite{law}. In this paper, we extend this method and present a validation study using discrete-structure simulations of an artificial cantilever-rod. In the validation study, we consider a cantilever rod with an artificially constructed, discrete, lattice-structure. Large deformations in two dimensions are then simulated. An effective constitutive-law is estimated from these deformations using inverse methods. Finally, we test the estimated law by employing it in the continuum rod and comparing the simulation results with those of discrete-structure simulations under a variety of cantilever loading-conditions. Once validated, the method can then leverage MD simulations of the discrete atomistic structure of DNA and other long-chain bio-molecules to estimate their constitutive laws.

We begin by first reviewing the mathematical formulation of the forward rod-model~\cite{goyal:05b} in Section~\ref{rod model} and then introduce a general, conceivable form of the constitutive law. Then in Section~\ref{lattice}, we describe the cantilever example and its validation plan for the estimation of the constitutive law from some measurable outputs. We close by summarizing the conclusions.

\section{The Computational Rod Model}
\label{rod model}

\begin{figure}[h!]
 \centering
 \includegraphics{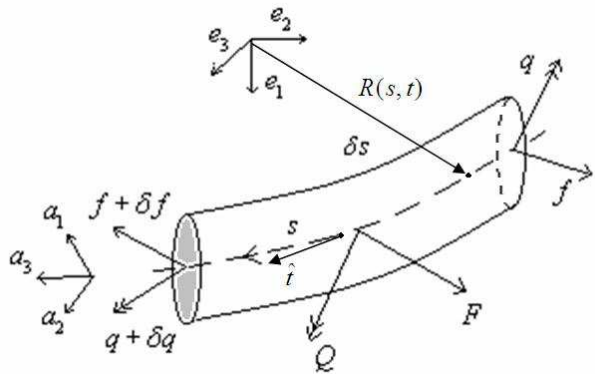}
  \caption{Dynamical rod model of double-stranded DNA on long-length scales. Helical axis of duplex defines the rod centerline which forms a three-dimensional space curve located by $\vec{R}\left(s,t\right)$.}
 \label{fig:dnarod}
 \end{figure}

In this section, we review the mathematical formulation of the forward rod model~\cite{goyal:05b}. The dynamics of rod deformation follow from the rigid-body motion of its individual cross-sections shown in Figure~\ref{fig:dnarod}. To track the rigid body motion of each cross-section, we fix at its mass center a reference frame  $a{_i}(s,t)$, where $s$ is the (unstretched) centerline coordinate and $t$ is time. The rigid body motion of the cross-section is described by its translational velocity $\vec{v}(s,t)$ and its angular velocity $\vec{\omega}(s,t)$. The gradient of these two vector fields along the rod's centerline captures the rate of change of the rod's deformation. The curvature vector, $\vec{\kappa}(s,t)$, describes this deformation of the rod's centerline (helical axis). The undeformed state is denoted by $\vec{\kappa_{0}}(s)$. Rod deformation in general involves bending curvature about two axes, shear along two axes, torsion, and extension (or compression). The position of the cross-section reference frame is denoted as $\vec{R}(s,t)$. The gradient of $\vec{R}(s,t)$ along $s$ is given by $\vec{r}$ and points along the centerline tangent $\hat{t}(s,t)$. In a stress-free state $\vec{r} = \hat{t}$.  The stress distribution across the cross-section results in a net internal (tensile and shear) force $\vec{f}(s,t)$ and (bending and torsional) moment $\vec{q}(s,t)$.

The governing equations for the rod dynamics are~\cite{goyal:05b}:
\begin{eqnarray}
\frac{\partial \vec{v}}{\partial s} + \vec{\kappa} \times \vec{v} &=& \frac{\partial \vec{r}}{\partial t} + \vec{\omega} \times \vec{r}, \label{position_continuity} \\
\frac{\partial \vec{\omega}}{\partial s} + \vec{\kappa} \times \vec{\omega} &=& \frac{\partial \vec{\kappa}}{\partial t}, \label{orient_continuity} \\
\frac{\partial \vec{f}}{\partial s} + \vec{\kappa} \times \vec{f} &=& m\frac{\partial \vec{v}}{\partial t} + m \vec{\omega} \times \vec{v} - \vec{F}, \label{linear_momentum} \\
\frac{\partial \vec{q}}{\partial s} + \vec{\kappa} \times \vec{q} &=& \textrm{I}\frac{\partial \vec{\omega}}{\partial t} + \vec{\omega} \times \textrm{I}\vec{\omega} + \vec{f} \times \vec{r} - \vec{Q}, \label{angular_momentum}
\end{eqnarray}
where $m(s)$ denotes the mass of the rod per unit length and the tensor $\textrm{I}(s)$ denotes mass moments of inertia per unit length,  $\vec{F}(s,t)$ is the distributed external force per unit length and  $\vec{Q}(s,t)$ is the distributed external moment per unit length. The partial derivatives are all relative to the cross-section-fixed reference frame $\vec{a_{i}}(s,t)$. (\ref{position_continuity}) follows from the continuity of $\vec{R}(s,t)$. (\ref{orient_continuity}) follows from the continuity of the cross-section orientation by constraining the curvature (twist) and angular velocity vectors. (\ref{linear_momentum}) and (\ref{angular_momentum}) are the Newton-Euler equations for an infinitesimal rod element. 

The stress-strain relationship in the rod integrated over its cross section may result in equations of the form 
\begin{equation}
\psi_{i} \left( \vec{\kappa}, \vec{q}, \vec{f}, s \right) = 0 . 
\label{const}
\end{equation}

(\ref{const}) represents the general form of a nonlinear, elastic constitutive-law. The constitutive law of the rod follows from its interatomic interactions. Therefore the varying sequence of base-pairs along the length of DNA results in the explicit dependence of $\psi_{i}$ on $s$. The result is a non-homogeneous constitutive-law, as captured in (\ref{const}). The basic form of this constitutive law and its sequence-dependent mapping is an open area of research. Current efforts to determine the sequence-dependence of the constitutive law always begin by assuming a linear approximation to (\ref{const})~\cite{beveridge:04a,dixit:05a}. It is important to note, however, that the assumption of linearity for the material law of DNA has been recently questioned and debated~\cite{cloutier:04a,du,wiggins:05a}, wherein the kink-ability of the molecule indicate non-convexity in the constitutive law.

\section{Constitutive Law Estimation from a Lattice-Structure}
\label{lattice}

\begin{figure}[h!]
\centering
\includegraphics{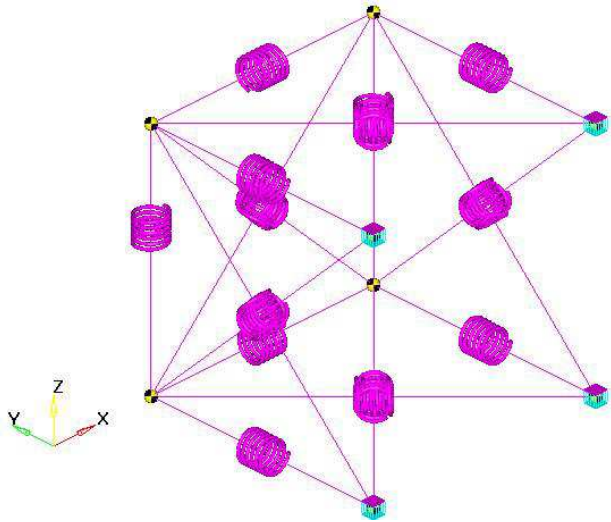}
\caption{A unit cell of the discrete-lattice model of DNA. Eight point masses are displaced a unit distance from their nearest-neighbors to form a cube. Springs of unit linear stiffness are then attached to each mass forming an unstretched configuration. The first unit cell, shown here, has the face of the cube without springs fixed to resemble the clamp of a cantilever beam.}
\label{unitlattice}
\end{figure}

In this section, we consider a discrete-lattice structure of a cantilever in static equilibrium, which is fixed at one end and loadings are applied to the free end. We construct the discrete-lattice model using the commercial, multi-body-dynamics software, Hyperworks. We followed the default system of units in Hyperworks, which is [Newton, Millimeter, Kilogram, Second]. Figure~\ref{unitlattice} shows the unit lattice cell where point masses are inter-connected by linear springs with a spring constant of unity. The point masses are arranged to form a unit cube where each mass is one unit of distance away from its nearest neighbors. The springs are connected along each edge of the cube and also along the diagonal of each face. The cubic cell is repeated thirty times to serve as the lumped-parameter model of a cantilever.

For the static equilibrium of the cantilever, the continuum-rod formulation described in Section~\ref{rod model} reduces to a time-independent formulation where all partial derivatives with respect to $t$ vanish and $\vec{v} = \vec{\omega} = 0$ as well. The equations of motion simplify to two ordinary vector differential equations in the spatial domain, $s$ where,
\begin{eqnarray}
\frac{d\vec{q}}{ds} + \vec{\kappa} \times \vec{q} &=& \vec{f} \times \vec{r}, \\
\frac{d\vec{f}}{ds} + \vec{\kappa} \times \vec{f} &=& 0.
\label{simple}
\end{eqnarray}

The direction of increasing $s$ is from the clamped end to the free end. By the symmetry of the lattice-structure, the constitutive law is decoupled in the principle directions of bending and torsion and does not have any coupling with force $\vec{f}(s,t)$. Choosing a body-fixed frame along the principal directions, (\ref{const}) simplifies to three scalar relations,
\begin{equation}
\psi_{i} \left( \kappa_{i}, q_{i} \right) = 0, 
\end{equation}  
where $i = 1, 2, $ or $3$ corresponding to the principal axes, $a_{i}$. Specifically, we let  $a_{1}$,  $a_{2}$, and  $a_{3}$ correspond to the first bending, second bending, and torsion axes, respectively.

With this decoupling in place and the assumptions of inextensibility and unshearability~\cite{goyal:05b}, we can further simplify our study by looking at deformations about one principal axis at a time. Hence we focus our further explanation on planar bending. For the case of a shear force applied in the $a_{1}$-direction in-plane bending occurs about the $a_{2}$-axis. The first and second components of $\vec{\kappa}$ and $\vec{q}$ and the second component of $\vec{f}$ vanish. In this case the governing equations of the rod model further reduce to the following three nonlinear ODEs,
\begin{eqnarray}
\frac{dq_{2}}{ds} = -f_{1} , \label{gov1}\\
\frac{df_{1}}{ds} = -f_{3} \kappa_{2}, \label{gov2}\\
\frac{df_{3}}{ds} = f_{1} \kappa_{2}  \label{gov3}
\end{eqnarray}
and the simplified constitutive law, 
\begin{eqnarray}
\psi_{2} \left( \kappa_{2}, q_{2} \right) =0. \label{law}
\end{eqnarray}
(\ref{gov1}), (\ref{gov2}), and (\ref{gov3}) can be rewritten further in state-space form by defining the state vector, $x(s)$, where $x(s) = \left[ q_{2}(s), f_{1}(s), f_{3}(s) \right]^{T} \in \mathbb{R}^{3}$ and input $u(s) = \kappa_{2}$. A causal relationship need not exist between input $u$ and state $x$. The above state-space equations are nonlinear with an unknown algebraic constraint, (\ref{law}). We note that the state vector $x(s)$ represents the bending moment, shear force, and tension. Similarly, the input $u(s)$ represents the curvature. We prescribe the state vector at the free end of the cantilever, i.e. at the final value of $s$. We convert the constrained final value problem above into a constrained initial value problem by substituting $s = -s'$ so that

\begin{eqnarray}
\frac{dq_{2}}{ds'} = f_{1} , \label{gov4}\\
\frac{df_{1}}{ds'} = f_{3} \kappa_{2}, \label{gov5}\\
\frac{df_{3}}{ds'} = -f_{1} \kappa_{2}  \label{gov6}
\end{eqnarray}

In terms of the state vector $x(s')$ and input $u(s')$ the above equations are of the form
 
\begin{eqnarray}
\frac{dx}{ds'} = f(x,u) \label{statespace}
\end{eqnarray}

with the constitutive law (\ref{law}) as an unknown constraint.

\begin{eqnarray}
\phi(x,u)=0 \label{constraint}
\end{eqnarray}

The above state-space equations can be solved as an initial value problem employing an appropriate DAE integrator if the constitutive law is known. We use MATLAB Simulink to set up the above equations. By contrast, for the estimation of the constitutive law from some measurable data, we treat the inverse method as a state-estimation and input-reconstruction problem. For this purpose we employ the techniques proposed in~\cite{law} which are based on filtering and input estimation-algorithms~\cite{palanthUMVACC2007,palanthwindACC2008,steven_kitanidis}. 

In the next section, we illustrate a validation scheme for the estimation method in~\cite{law}. In particular, we assume that the only measurements available from the discrete structure simulations are the shape of the deformed cantilever and the loads at the free end. From the shape, we can compute the curvature $\kappa_{2}(s')$, which can then be fed as a known input in the ODE (\ref{statespace}) to solve for the state $q_{2}(s')$. To estimate the functional relationship (\ref{law}) between $\kappa_{2}$ and $q_{2}$ we express the function, $\psi_{2}$, as a sinusoidal-basis-function expansion, where the unknown coefficients of this expansion are then found by standard least-squares-fitting. 

Once (\ref{law}) has been estimated for a certain set of loading conditions on the discrete lattice-structure, we numerically solve the governing equations given by (\ref{statespace}) for the continuum rod-model, employing this estimated constitutive-law. Finally, to validate the estimated constitutive-law, we apply different loading conditions to the discrete structure and compare these results with corresponding simulations of the continuum rod-model using the estimated constitutive-law. The results of these comparisons determine whether the estimated constitutive-law captures the mechanical behavior of the discrete structure in its continuum rod-model approximation. 

\section{Results}
\label{results}

\begin{figure}[h!]
\centering
\includegraphics{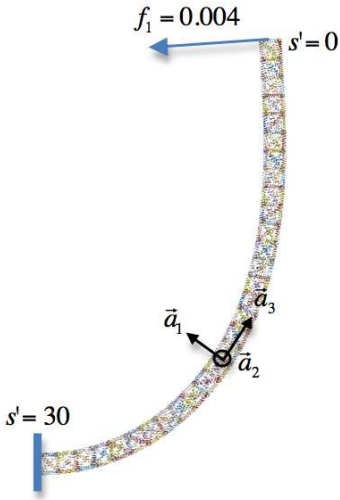}
\caption{Static equilibrium shape of the discrete-structure cantilever. Case 1: prescribed shear force, but no tension and no bending moment at the free end.}
\label{case1}
\end{figure}

\begin{figure}[h!]
\centering
\hspace{-0.1in}\includegraphics[width=60mm]{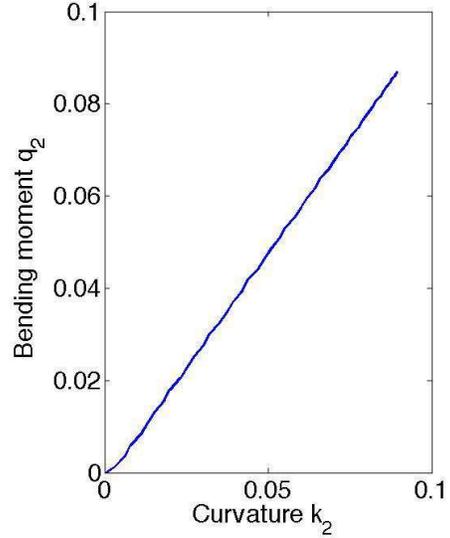}
\caption{Estimated constitutive law from the case 1 simulation of the discrete structure.}
\label{ConstitutiveLaw}
\end{figure}

We begin by simulating the static equilibrium configuration of the discrete, cantilever structure with a shear force $f_{1}=0.004$ as shown in Figure~\ref{case1} using the Force Imbalance Method-Type D for the static analysis in Hyperworks. The system of units was set to the default in Hyperworks, which is [Newton, Millimeter, Kilogram, Second]. From this configuration data, we calculate the curvature $\kappa_{2}(s')$ to use as an input to the Simulink model of the continuum rod described in the Section~\ref{rod model}. The Simulink model calculates the bending moment $q_{2}(s')$ according to the continuum-rod formulation. Based on these computations, Figure~\ref{ConstitutiveLaw} shows the estimated constitutive relationship between the bending moment and curvature.

\begin{figure}[h!]
\centering
\hspace{-0.1in}\includegraphics[width=60mm]{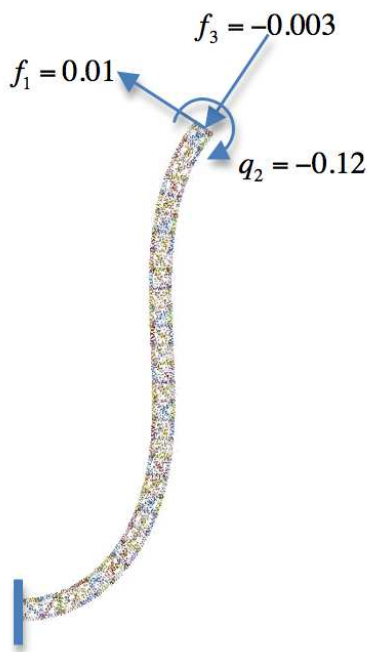}
\caption{Static equilibrium shape of the discrete-structure cantilever. Case 2: prescribed shear force, tension and bending moment at the free end.}
\label{case2}
\end{figure}

\begin{figure}[h!]
\centering
\hspace{-0.1in}\includegraphics[width=60mm]{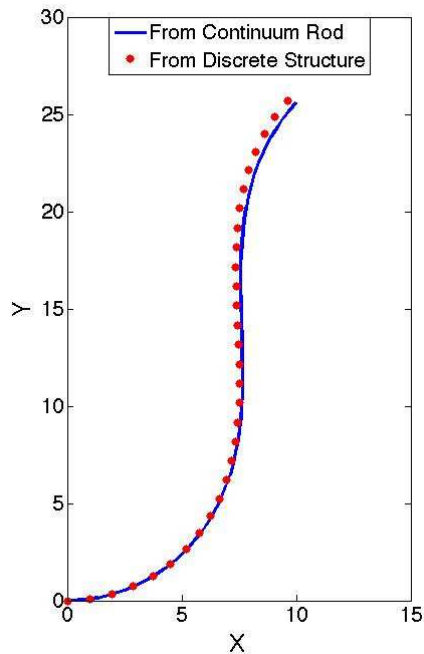}
\caption{Centerline in case 2 loading predicted from discrete-structure simulation (dots) and continuum-rod simulation (solid curve) employing the estimated constitutive-law.}
\label{comparison}
\end{figure}

In order to validate the estimated constitutive-law, we test it in several different loading environments and evaluate if it predicts the same deformation from the rod formulation as it does from the discrete-structure simulations. Here we present on such validation test. The loading conditions are shown for the validation test in Figure~\ref{case2} on the deformed equilibrium of the discrete structure. In this case we apply shear force, compressive force and bending moment at the free end. We apply these loading conditions in the Simulink model of the continuum rod employing the estimated constitutive-law shown in Figure~\ref{ConstitutiveLaw}. Figure~\ref{comparison} shows the comparison of the centerline predicted from discrete-structure simulation (dots) and the continuum-rod simulation (solid curve) employing the estimated constitutive-law. The closely matching shape validates the estimated constitutive-law.

It is likely that the nearly linear result of the estimated constitutive-law arises from the fact that the diagonal springs of the discrete structure are only negligibly deformed compared to those parallel to the length of the structure. 
A nonlinear law would be expected if deformations of the structure were simulated such that the diagonal springs experienced large deformations.   


The close agreement of the centerline deformations of the discrete-lattice structure and the continuum rod-model also suggests that a discrete model may be an appropriate method to determine a constitutive law for DNA. The agreement is noteworthy when considering that each simulation is run independently with different software and solvers.  We also note that, while inextensibility and unshearability are assumed in the forward rod-model, no such assumptions were imposed on the discrete-lattice cantilever. The slenderness ratio (length/thickness) of the discrete structure is 30. Therefore the results suggest a confirmation of the assumptions of Kirchhoff Rod Theory.


\section{Conclusions}
\label{conclusions}

Elastic rod models have recently emerged as efficient tools to simulate long length and time scale deformations of DNA which are crucial to its biological functions including gene expression. However, the usefulness of the rod model to simulate DNA is severely limited by the lack of knowledge of an accurate constitutive law for the molecule, which in general, is expected to be nonlinear. Discrete structure MD simulations of the molecule for small length and time scales can provide ample data for the estimation of an accurate, constitutive law. However, the past efforts in mapping the constitutive law from MD simulations have been limited to linear approximations or a few parameter estimations. The challenge has been to develop inverse methods to estimate the functional form of the constitutive law. 

In this paper, we presented an estimation method and a validation study using discrete-structure simulations for a simple, but relevant, example motivated by modeling DNA. We considered a cantilever rod with an artificially constructed, discrete, lattice-structure. Large deformations in two dimensions are then simulated. An effective constitutive law is estimated from these deformations using inverse methods. Finally, we tested the estimated law by employing it in the continuum rod model and comparing the simulation results with those of discrete-structure simulations under new, cantilever loading-conditions. We believe this method can then leverage MD simulations of the discrete atomistic structure of DNA and other long-chain bio-molecules to estimate their constitutive laws.

\section{Acknowledgments}
\label{acknow}

The authors gratefully acknowledge Altair Eng., Inc. for providing their CAE tool Hyperworks for this research. The authors Adam R. Hinkle and Sachin Goyal also gratefully acknowledge the training provided by Mr. Sundar Nadimpalli, Application Specialist - Multibody Dynamics, Altair Eng., Inc., Bangalore to construct and simulate a lattice structure model in Hyperworks.

\bibliographystyle{asmems4}
\bibliography{paper}

\end{document}